\documentclass{ws-gm}
\usepackage{xcolor}
\usepackage[spanish]{babel}
\usepackage[utf8]{inputenc} 
\usepackage[T1]{fontenc}
\usepackage{lmodern}
\usepackage[verbose]{hyperref}
\def\xa{x^{(a)}}

\def\va{v^{(a)}}

\def\Aa{A^{(a)}}

\def\Pa{P^{(a)}}

\def\Ja{J^{(a)}}

\def\Ha{H^{(a)}}

\def\Ga{G^{(a)}}

\def\Ka{K^{(a)}}

\def\Xa{X^{(a)}}

\def\px#1#2{\partial_{x_{#2}^{(#1)}}}
\def\pv#1#2{\partial_{v_{#2}^{(#1)}}}
\def\dis{\displaystyle}
\def\pref#1{(\ref{#1})}
\def\NR{\mathbb{R}}
\mathchardef\mhyp="2D 
\renewcommand*{\thefootnote}{\arabic{footnote}}

\begin{document}
\renewcommand{\bibname}{References}

\def\wsdraftnote{\quad\qquad\qquad}%

%
\catchline{}{}{}{}{}
%

\title{Interaction anomalies and one-particle dynamics\\ in very special relativity theories\\
}
\renewcommand*{\thefootnote}{\arabic{footnote}}

\author{Manuel Asorey\footnote{asorey@unizar.es}  , Fernando Falceto
  \footnote{
falceto@unizar.es    } }

\address{Centro de Astropart\'iculas y F\'isica de Altas Energ\'ias, Departamento de F\'isica Te\'orica, Universidad de Zaragoza, E-50009 Zaragoza, Spain\
}
\author{Giuseppe Marmo \footnote{marmo@na.infn.it}}

\address{Dipartimento di Fisica “E. Pancini”, Università di Napoli Federico II and INFN-Sezione di Napoli, Complesso Universitario di Monte S. Angelo Edificio 6, via Cintia, 80126 Napoli, Italy, 
}

\maketitle


\begin{abstract}
It is well known that relativistic invariance introduce
strong  constraints in the interactions of classical particles. 
We generalize  the non-interaction theorems for Lorentz violating systems which  still  preserve a 
subgroup of Poincaré symmetry. In particular we analize the case of very special 
relativity introduced by Cohen and Glashow. We also extend the analysis for Galilei invariant 
multiparticle systems and  for some anisotropic systems which are still invariant under some 
maximal subgroups of Galilei group.
\end{abstract}

\keywords{No-interaction theorems; Very special relativity; World line conditions.}

\ccode{Mathematics Subject Classification 2020:  83A05, 70S10, 70H40}

\section {Introduction}

The no-interaction theorems developed in the sixties of last century pointed out the
difficulties of  introducing relativistic interactions for point-like classical  particles.
Usually, this theorem is used to justify that relativistic interactions can only be
properly implemented by means of fields in  field theories. The difference being basically
that in such a case interaction is only possible thanks to the presence of an infinity number of 
degrees of freedom.

Although the original formulation of the non-interaction theorem was given in the canonical
formalism \cite{basic} (see also \cite{basic1,cannjord,leu, sudars, sudmuk, sudmuk1,sudmuk2}) it is also possible to obtain a proof in the Lagrangian formalism
\cite{BalMarStern,Lagrangian}. In fact, the theorems can also be proven by using the world line
conditions (WLC)  to define the  Poincaré symmetry in terms of differential operators in the tangent bundle
$TR^{3N}$ of the configuration space $R^{3N}$  of a family of $N$ particles \cite{CCIM}. The anomalies
 found in the commutators of such  generators imply the vanishing of the interaction between the
 $N$-particles.

%
%
%
%

In the last years the search of Lorentz invariance violating (LIV) effects has being intensified
by the  observation of the behavior of high energy cosmic rays \cite{QGP}.
Among the theories which could explain slight departures of 
Poincaré invariance   stand out the theories known 
as very special relativity  theories (VSR). These theories break Lorentz invariance but
still  preserve a subgroup of Poincaré symmetry \cite{CG}. See the recent
thesis \cite{AS} for a broad view on phenomenology.  In this paper we analyze the
possible existence of non-trivial interactions in VSR theories trying to escape the constraints
leading to non-interaction theorems.  We  extend the analysis to other LIV theories 
preserving other subgroups of Poincaré group. 

We also generalize the analysis of non-interaction  theorems for non-relativistic theories which
are invariant under Galilei group \cite{Gupta} or some of its subgroups.

The organization of  the paper is as follows. In Section 2 we introduce the basic notion of
world line conditions  (WLC) that translate the space-time symmetries to a fixed time Cauchy data
framework. Next we introduce in Section 3 the more special relativity symmetries  both for
 Galilei and Poincaré symmetry breaking relativity theories and analyze the role of these symmetries
 in the proofs of non-interaction theorems. Finally, we consider the case of 
 homogeneous symmetry subgroups  which do not include the space-time translations in Section 4.

\section{The World Line Conditions (WLC)}

In this section we introduce the generators of the Galilei and Poincaré
algebra as they act in the space of trajectories by means of the so called WLC.
Although in most of the applications we will restrict to the case of
a single particle, in order to give a more general perspective we shall deal in this
section the case of several particles labeled by superindices $(a),\ (b),\dots$.
The corresponding positions and velocities at a fixed time are denoted
by $\xa=(\xa_i)$ and $\va=(\va_i)$. 

We assume the dynamics is given, i. e. we know the acceleration ${\dot v}^{(a)}$
of every particle as a function of positions and velocities, it will be
denoted by $\Aa(v,x)$, where $v=(\va)$ and $x=(\xa)$ stand for velocities and
positions of all the particles.

Next we would write the generators of the corresponding
relativity group as vector fields acting on the world-lines
(solutions of the equations of motion) parametrized by its position and
velocity at a given time.

Except for the boosts, that differ for Galilei and Poincaré relativity
groups, the other generators are common, they can be written
  \begin{alignat*}{3}
    P_i&=\sum_a \Pa_i&\quad\quad  \mbox{with\quad\quad} &\Pa_i=\px ai\\
    J_i&=\sum_a \Ja_i&\quad\quad  \mbox{with\quad\quad}
    &\Ja_i=\epsilon^{ijk}\left(\xa_j\px ak+\va_j\pv ak\right)
    \\
       H&=\sum_a \Ha&\quad\quad  \mbox{with\quad\quad}
    &\Ha= \va_i\px ai+\Aa_i(v,x)\pv ai,
  \end{alignat*}
  where $H$ stands for time translation of the trajectory and the implicit
  sum of repeated spatial indices, $i,\ j,\ k,\ l$, is assumed.

  Remark: The reader should appreciate that unlike the usual realization
  of the Poincaré group, where time translation is simply given by $\partial_t$,
  i.e., purely kinematical, here time translation is dynamically defined
  and this is the essence of the world-line-conditions, the dynamical
  evolution is required to represent time translation in the realization
  of the Poincaré group. This dynamical ingredient requires that the Poincaré
  generators are realized on the tangent bundle in a specific way that
  depends on the dynamics.

  As already mentioned, boosts behave differently for Galilei and
  Ponciaré relativity groups. They will be named respectively $G_i$
  and $K_i$ and are given by
  \begin{alignat}{3}\label{equ:boosts}
    G_i&=\sum_a \Ga_i&\  \mbox{with\ } &\Ga_i=-\pv ai\\
    K_i&=\sum_a \Ka_i&\ \mbox{with\ }
    &\Ka_i=\xa_i\va_j\px aj+\left(\va_i\va_j-\delta_{ij}+\xa_i \Aa_j(v,x)\right)
    \pv aj\nonumber.
  \end{alignat}

  The question now is under what circumstances, i.e. for which form
  of the accelerations, the above generators fulfill the Galilei or Poincaré
  algebras?.

  We first compute the commutation relations of the vector fields
  associated to the common generators, $P_i,\ J_i,\ H$. It is immediate to get
  \begin{equation}\label{equ:PJcomm}
    [P_i,P_j]=0,\qquad[J_i, P_j]=-\epsilon^{ijk}P_k,\qquad
    [J_i, J_j]=-\epsilon^{ijk}J_k,
    \end{equation}
 wheras the commutators with $H$ happen to be
  \begin{equation}\label{equ:Hcomm}
    [P_i,H]=\sum_a P_i\Aa_l\ \pv al,\quad
  [J_i,H]=\sum_a\left(J_i \Aa_l-\epsilon^{ijl} \Aa_j\right)\pv al,
  \end{equation}
  that pick up an anomalous piece depending on the accelerations.
  
  But both commutators should vanish if they have to provide  a realization
  of the Galilei or the Poincaré algebra. Thus we get the first 
  two conditions for fulfilling the relations of any of the two relativity
  groups.
\vskip 3mm
  {\bf Condition I:}\quad
  $\dis P_i \Aa_l(v,x)\equiv\sum_b \px bi \Aa_l(v,x)=0,\ \mbox{for any } i\ ,l\ , a$
\vskip 1mm
  Which means that the acceleration must be translation invariant.

  From the second equation in \pref{equ:Hcomm} we get
\vskip 3mm
  {\bf Condition II:}\quad
  $\dis J_i \Aa_l(v,x)=\epsilon^{ijl} \Aa_j(v,x),\ \mbox{for any } i\ ,l\ , a$
  \vskip 1mm
  Which means that the acceleration should transform as a vector under global
  spatial rotations.

  The two previous conditions look natural if one wants to have dynamics
  with relativistic symmetry group that includes space-time
  translations and rotations. Let us consider which conditions are obtained when
  we add boosts.

  In this case we must distinguish between the two possibilities Galilei and
  Poincaré relativity gropus. We start with the first one.
  \vskip 3mm
  \subsection{Galilei relativity group}

  In thís case the boosts are generated by $G_i$ as defined in
  \pref{equ:boosts} and their commutation relations
  \begin{align}\label{equ:GalHcomm}
[H,G_i]&=P_i+\sum_a G_i\Aa_l(v,x)\pv al\\
    [P_i,G_j]&=0,\quad[J_i,G_j]=-\epsilon^{ijk}G_k,\quad [G_i,G_j]=0,\label{equ:homGalcomm}
  \end{align}
  correspond to those of  Galilei algebra without central extension%
  \footnote{If we add a central extension, with central charge $M^{(a)}$ for every
  particle, the anomalous terms remain unchanged. Hence, in order to keep the
  expressions simpler, we limit ourselves to the case without central extension.},  
  provided the following condition is  satisfied.
  \vskip3mm
 {\bf Condition IIIG:}\quad
 $\dis  G_i \Aa_l(v,x)=0,\ \mbox{for any } i\ ,l\ , a.$
 \vskip 1mm
The physical meaning of such a condition is that the 
accelerations must be invariant under the simultaneos translations
 of all velocities.

 Now it is easy to see that for one single particle the only
 dynamics compatible with the Galilei group is the free one with vanishing
 acceleration.

 The proof is simple. Conditions I and IIIG impose
$$P_i A_l(v,x)=0,\ \ G_i A_l(v,x)=0,$$
 where we have removed the superindices as we have only one particle,
 therefore $A_l$ is constant but then $J_jA_l=0$. Therefore
 from condition II we get
 $$\epsilon^{ijl} A_j=0$$
 implying that the acceleration must vanish.

  This non-interaction theorem is not valid any more for several Galilean
 particles where vectorial accelerations depending only on the relative
 positions and velocities fulfill all three conditions.
 
 For instance, for two particles, the most general form of the accelerations
 satisfying all three conditions is parametrized by real functions defined
 on $\NR^3$, $f^{(a)}$ $g^{(a)}$, $a=1,2$, and given by
 \begin{align*}
 \Aa_l&=
 v^{(1\mhyp2)}_lf^{(a)}\big(\,(v^{(1\mhyp2)})^2,\; 
 v^{(1\mhyp2)}\!\cdot x^{(1\mhyp2)},\;(x^{(1\mhyp2)})^2\big)\\[2mm]
 &+
x^{(1\mhyp2)}_lg^{(a)}\left(\,(v^{(1\mhyp2)})^2,\;
 v^{(1\mhyp2)}\!\cdot x^{(1\mhyp2)},\;(x^{(1\mhyp2)})^2\right),
 \end{align*}
 where $v^{(1\mhyp2)}_l$ and $x^{(1\mhyp2)}_l$ denote respectively
 $v^{(1)}_l-v^{(2)}_l$ and $x^{(1)}_l-x^{(2)}_l$ while the arguments
 of the functions are their modulus square and their scalar product.

 \subsection{Poincaré relativity group}
 The boosts defined in the second line of eqs. \pref{equ:boosts}
 that correspond to the Poincaré algebra satisfy.
   \begin{align}
     [\,P_i,K_j]&=\delta_{ij}H+\sum_a\xa_j\, P_i\Aa_l\,\pv al,\label{equ:anopoin1}\\
     [\,J_i,K_j]&=-\epsilon^{ijk}K_k + \sum_a \xa_j\left(J_i\Aa_l-\epsilon^{ikl}\Aa_k\right)\pv al,\label{equ:anopoin2}\\
     [\,H\,,K_i]&=P_i+\sum_a \left(\Xa_i\Aa_l+ \va_l \Aa_i\right)\pv al,\label{equ:anopoin3}\\
     [K_i,K_j]&=\epsilon^{ijk}J_k+\sum_a
     \left[\,\xa_i\left(\Xa_j\Aa_l+ \va_l \Aa_j\right) -\ i\leftrightarrow j\ \right]\,\pv al\label{equ:anopoin4},
  \end{align}
   where $$\Xa_i = 2\va_i + \xa_iH - K_i.$$

The Poincaré algebra is fulfilled if all the anomalous terms depending
on the accelerations in the above equations  vanish. Because the anomalies in \pref{equ:anopoin1} and \pref{equ:anopoin2}
cancel due to conditions I and II
this imposses only a new requirement.
\vskip3mm
{\bf Condition IIIP.}\quad$\dis \Xa_i\Aa_l+ \va_l \Aa_i=0, \mbox{\ for all\ } i,\,l,\,a.$
\vskip 1mm
\noindent  Because condition IIIP guarantees that 
\pref{equ:anopoin3} and \pref{equ:anopoin4} vanish.

From now on we can repeat the analysis carried out for the Galilei relativity group.
If we consider one single particle and, consequently, we remove the superindices,
conditions I and II imply
$$A_l(v,x)=v_lf(v^2).$$

On the other hand, using the definitions
$$X_i=2v_i+(\delta_{ij}- v_i v_j) \partial_{{v_j}}$$
and from the condition
$$X_iA_l+v_l A_i=0$$
we deduce that $f$ has to satisfy
$$v_iv_l(f+2(1-v^2)f')+\delta_{il}f=0,\ \mbox{for}\  i, l=1,2,3.$$
Taking $i=1,\ l=2$ we see that the parenthesis vanishes
(the alternative, $v_1$ or $v_2=0$, does not change the conclusion). 
Hence, from the equation with $i=l=1$ we deduce that $f=0$. 

We have seen that, similarly to the Galilean case, the dynamics
of a single particle may give rise to the Poincaré algebra only if it moves freely.
The multiparticle case is left for a further research.

Let us now move into  a different direction.
We keep the one particle case but we reduce the relativity goup by taking
different subgroups to see if the non-interaction theorem
(free motion of the particle) also holds in this less restrictive scenario.

\section{More special relativity groups}

As already mentioned, the strategy for trying to find non trivial dynamics in the single particle
case will be to break some of the symmetries of the relativity group.

In order to clarify the situation it is convenient to study the most restrictive cases
i. e. the maximal proper subgroups or subalgebras. This is what we shall denote as more special
relativity groups.

Notice that the Abelian subalgebra generated by $\{P_i,\, H\}$ is an ideal,
both in Galilei and Poincaré algebras, therefore it should be contained
in any more special relativity group, (with the only exception of the homogeneous
subalgebra, complementary to it).

We shall study separately the Galilei and Poincaré cases starting with
the first one.

\subsection{More special Galilei groups}

It is easy to see that we have three possible maximal propers subgroups.
They will be called static, anisotropic and
very special subgroups.

\subsubsection{Static subgroup}

We call it static because the relative motion of observers is not allowed,
i. e. all boosts are removed and its subalgebra is generated by
$$\{P_i,\,H,\, J_i\}.$$
The only anomaly cancelations we should care about are
$$P_i A_l(v,x)=0,\quad J_i A_l(v.x)-\epsilon^{ijl}A_j(v,x)=0$$
The general solution of this vanishing condition is given by
$$A_l(v,x)=v_l f(v^2).$$

\subsubsection{Very special subgroup}
The name is taken from the subgroup of the Poincaré algebra, that
actually motivates our work, introduced in \cite{CG}. It somehow
interpolates between the previous two. The maximal subalgebra is
$$\{P_i,\, H,\,\beta G_1-J_2,\,\beta G_2+J_1,\,J_3\},$$
where $\beta$ is a constant, introduced by dimensional reasons, that
interpolates between the static and anisotropic subalgebras.

We first consider the anomalies induced by $P_i,\ J_3$ which restricts
the accelerations to
\begin{equation}\label{equ:vsacc}
  A_\alpha(v,x)=v_\alpha f(\vec v^2,v_3),\ \alpha=1,2
  \quad A_3(v,x)=g(\vec v^2,v_3).
\end{equation}
where $\vec v^2=v_1^2+v_2^2$.

To cancel the anomalies of the other two generators of the algebra
we have to satisfy the conditions
\begin{align}
  &(G_iA_l-J_2A_l+\epsilon^{2jl}A_j=0\label{equ:vsanom1}
  \\
  &(G_2A_l+J_1A_l+\epsilon^{1jl}A_j=0.\label{equ:vsanom2}
\end{align}
Now, using the ansatz in \pref{equ:vsacc} and
the form of the operators associated to $G_i$, $J_i$,
the above equations  translate  into differential
equations for $f$ and $g$.

Consider first the equations \pref{equ:vsanom1} for $l=1,2,3$
\begin{align*}
  l=1) \quad &v_1((\beta-v_3)2v_1f'+v_1\dot f)+(\beta-v_3)f+g=0\\
  l=2) \quad &v_2((\beta-v_3)2v_1f'+v_1\dot f)=0\\
  l=3) \quad &(\beta-v_3)2v_1g'+v_1\dot g-v_1f=0
\end{align*}
where $f',\ g'$ stands for the derivative of $f$ and $g$ with respect to their
first variable and $\dot f,\ \dot g$ with respect to the second.

Smilarly from \pref{equ:vsanom2} we derive
\begin{align*}
l=1) \quad &v_1((\beta-v_3)2v_2f'+v_2\dot f)=0\\  
l=2) \quad &v_2((\beta-v_3)2v_2f'+v_2\dot f)+(\beta-v_3)f+g=0\\  
l=3) \quad &(\beta-v_3)2v_2g'+v_2\dot g-v_2f=0
\end{align*}

It is easy to see that all six equations are solved if
$$2(v_3-\beta) f'-\dot f=0,\quad g=(v_3-\beta)f.$$
The solution of the differential equation is
$$f(\vec v^2,v_3)=F(\vec v^2+ (v_3-\beta))^2)$$
and the general form of the accelerations can be compactly written
as
$$A_l(v,x)=\partial_{v_l}W(\vec v^2+ (v_3-\beta)^2),\ l=1,2,3.$$
for any arbitrary function $W$.

Then, one checks that the result for the static subgroup is, as it should be, the
$\beta\to0$ limit of the very special case.

In the opposite limit, $\beta\to \infty$, which in practice means the removal of $J_1$
and $J_2$, this  is not a maximal proper subalgebra, as we can add $G_3$ to it.
The resulting subalgebra, that we study below, is the most special (highest dimensional)
of all subalgebras of the Galilei group.

\subsubsection{Most special (or anisotropic) subgroup}

The subgroup can be interpreted as the breaking of
the rotation invariance to the one-dimensional  subgroup of planar rotations around
the $x_3$ axis. Hence  the name anisotropic.

In this case the subalgebra is
$$\{P_i,\,H,\,G_i,\, J_3\},$$
and the possible anomalies
$$P_i A_l(v,x)=0,\quad G_i A_l(v,x)=0,\quad
J_3 A_l(v.x)-\epsilon^{3jl}A_j(v,x)=0.$$
It is not difficult
to show that the general solution is
a constant acceleration along the $x_3$ axis
$$A_1=A_2=0, A_3=g.$$
This is not very surprising, as the anisotropic subgroup is the group
of symmetries of a particle moving in a constant force field,
like the parabolic motion in uniform gravity.

\subsection{More special Poincaré groups}

We now turn our attention to the Poincaré group and its maximal proper
subgroups.

In these case we have a one parameter family of seven dimensional very special 
relativity groups and for some special value of the parameter an eight dimensional most special
group.

\subsubsection{Very special group}

Let us first consider the VSR algebra  generated by $\{P_i,\,H,\, K_1-\beta J_2,\, K_2+\beta J_1, J_3\}$.

After some work one can see that the most general form of the
accelerations compatible with the symmetry is
\begin{align*}
  A_\mu&=v_\mu(v_3-\beta)^2 F\left(\frac{v_3-\beta}{\sqrt{1-\vec v^2-v_3^2}}\right),
  \quad \mu=1,2\\
  A_3&=(v_3-\beta^{-1})(v_3-\beta)^2
  F\left(\frac{v_3-\beta}{\sqrt{1-\vec v^2-v_3^2}}\right)
\end{align*}

\subsubsection{Most special group}

The name comes from the fact that this is a special case of the very special
group, although it has some additional property.

For $\beta=1$ the previous algebra is not maximal any more.
It can be suplemented with $K_3$ giving rise to the highest dimensional proper
subalgebra of Poincaré. It is the analogous of the anisotropic Galilei
subalgebra. Both have dimension 8 and unique, up to scale, solution for
the acceleration.

The only solution is
\begin{align*}
  A_\mu&=g\,v_\mu\frac{\left(1-\vec v^2-v_3^2\right)^{3/2}}{v_3-1},
    \quad \mu=1,2\\
A_3&=g\left(1-\vec v^2-v_3^2\right)^{3/2}
\end{align*}
where $g$ is a constant.

\section{Homgeneous subgroups}

As we discussed before, the only maximal subgroups that do not contain
space-time translations are the homogeneous subgroups i. e. those
generated by rotations and boosts.

We will show that, while the more special relativity groups discussed
in previous section showed a clear parallelism between the Galilei and
Poincaré cases, in this occasion it is not like that.
In a way, the two relativity groups are at completely
opposite ends. This is the reason why we decided to treat them separately.

Let us first consider  the homogeneous Galilei subgroup. The commutators of
the generators of boosts and rotations determined by the WLC
are written in \pref{equ:PJcomm} and \pref{equ:homGalcomm}. We observe that they
do not have any anomalous term depending on the accelerations. This means that we do
not have any condition on them and the commutation relations of the
generators of the homogeneous Galilei group, induced by the WLC, fullfil the
Galilei algebra for any acceleration $A_l(v,x),\ l=1,2,3$.

As we already indicated, the Poincaré case, when we restrict to the Lorentz algebra, is at the opposite situation and the acceleration is bound to vanish. 
The proof of this result is as follows.

First, we should consider the conditions for cancellation of anomalies in
the commutators of the Lorentz algebra generated by
$\{J_i,K_i\}$. These can be read from eqs. \pref{equ:PJcomm},
\pref{equ:anopoin2} and  \pref{equ:anopoin4}. Therefore the conditions one has to
impose into the acceleration are
\begin{align}
  &J_i A_l-\epsilon^{ikl} A_k=0,\ i,l=1,2,3\label{equ:homPoiJ}\\
  &X_i A_l+ v_l A_i=0,\ i,l=1,2,3\label{equ:homPoiK}.
\end{align}

The general solution for eqs. \pref{equ:homPoiJ} is
$$A_l(v,x)=x_l\, f(v^2,v\cdot x,x^2)+v_l\, g(v^2,v\cdot x,x^2)$$
where $f, g:\NR^3\rightarrow \NR$ are real functions of three variables.
Next we will denote by $f_i, g_i$ their derivative with respect to
the $i$'th variable.

If we insert this solution into \pref{equ:homPoiK} we get
\begin{align*}
  X_i A_l+ v_l A_i&=  x_i x_l f_2+v_ix_l\left(2f+2(1-v^2)f_1-v\cdot xf_2\right)\\
  &+x_iv_l\left(f+g_2\right)
  +v_iv_l\left(2g+2(1-v^2)g_1-v\cdot x g_2\right)\\&+\delta_{il} g=0.
\end{align*}
This implies that each term must vanish separately wich implies $f=g=0$, i.e.
the acceleration must vanish $A_l(v,x)=0$

To summarize, we have shown that for the homogeneous Galilei
subgroups the acceleration is arbitrary while for the homogeneous
Poincaré it has to vanish. 

One interpretation of these two radically different results 
is that Galilei subgroup does not contain generators acting on time,
while the boosts of the homogeneous Poincaré  involve time shifts. It is just
this different characteristic of  both types of boosts that  allow the existence
or not of non-trivial interactions in relativistic systems 

\section*{Conclusions}

In summary,  the interacting dynamics of point-like particles
is very much constrained by relativistic principles. The constrains are weaker
in  Galilei relativity than in  Poincaré relativity.

We have simplified  our analysis restricting ourselves to the simplest case of one 
single particle but many of the arguments can be extended for multiparticle systems.
In particular, in Galilei relativistic systems we have shown that for one-particle systems
there is no possible interaction, whereas for multiparticle systems this is always possible.
In the Poincaré case the results are similar for one-particle systems, but we expect that
the non-interaction theorem  also holds for the multiparticle systems.

In   anisotropic VSR  relativistic theories  the anomalies arising in the
commutators of the infinitesimal generators of the symmetry algebra induce
weaker constrains on the type of interactions between point-like particles.
In particular, for static and VSR Galilei  invariant  one-particle systems it is
possible to have non-trivial interactions while preserving the invariance under
such a subgroups of Galilei group.
Similar results hold for VSR and most special relativity subgroups of Poincaré group,
where some interactions are compatible with the restricted  relativistic symmetries.

Finally,  in the case of  systems preserving 
 homogeneous symmetry subgroups  of Galilei or Poincaré groups,
  which do not include the space-time translations   we conclude that
  interaction is possible for the Galilei  case whereas it is not possible for the Poincaré case.
  This behavior  can be explained in physical terms because the Galilei boost  do not
  transform the physical time whereas Poincaré boost do.

%
%
%
%

\section*{Acknowledgements}

This paper is devoted to the memory of Miguel Muñoz Lecanda, who was one of the pioneers in Spain of the geometric approach to classical mechanics.  We would like to thank Miguel for friendship and for sharing his geometric and physical insights along the organization of 32 International Fall Workshops on  Geometry and Physics.
The work of M.A. and F.F. is partially supported  by Spanish MINECO/FEDER Grants No. PGC2022-126078NB-C21 funded by MCIN/AEI/ 10.13039/501100011033, ERDF A way of making Europe Grant; the Quantum Spain project of the QUANTUM ENIA of  {\sl Ministerio de Asuntos Econ\'omicos y Transformaci\'on Digital},  {\sl Diputaci\'on General de Arag\'on Fondo Social Europeo (DGA-FSE)} Grant No. 2020-E21-17R of  Aragon Government, and {\sl Plan de Recuperaci\'on, Transformaci\'on y Resiliencia}- supported European Union – NextGenerationEU Program on {\sl Astrof{\'{\i}}sica y F{\'{\i}}sica de Altas Energ{\'{\i}}as}, CEFCA-CAPA-ITAINNOVA. G. M. is  a member of the Gruppo Nazionale di Fisica Matematica (INDAM), Italy 


\end{document}